# Transforming Business with Generative AI: Research, Innovation, Market Deployment and Future Shifts in Business Models

Narotam Singh[a#], Vaibhav Chaudhary[b], Nimisha Singh[c], Neha Soni[d], Amita Kapoor[e#]

[a]Former Scientist, IMD, Ministry of Earth Sciences, Delhi, India. Email: narotam.singh@gmail.com
[b]Department of Mechanical Engineering, Indian Institute of Technology, Kharagpur, India
[c]School of Management, Bennett University, Greater Noida, UP, India,
[d]Department of Physics, MMEC, Maharishi Markandeshwar (Deemed to be University), Mullana-Ambala, Haryana, India,
[e]NePeur, India, Email: amita.kapoor@nepeur.com
[#] corresponding authors

## Abstract

This paper explores the transformative impact of Generative AI (GenAI) on the business landscape, examining its role in reshaping traditional business models, intensifying market competition, and fostering innovation. By applying the principles of Neo-Schumpeterian economics, the research analyses how GenAI is driving a new wave of "creative destruction," leading to the emergence of novel business paradigms and value propositions. The findings reveal that GenAI enhances operational efficiency, facilitates product and service innovation, and creates new revenue streams, positioning it as a powerful catalyst for substantial shifts in business structures and strategies. However, the deployment of GenAI also presents significant challenges, including ethical concerns, regulatory demands, and the risk of job displacement. By addressing the multifarious nature of GenAI, this paper provides valuable insights for business leaders, policymakers, and researchers, guiding them towards a balanced and responsible integration of this transformative technology. Ultimately, GenAI is not merely a technological advancement but a driver of profound change, heralding a future where creativity, efficiency, and growth are redefined.

## Introduction

Generative Artificial Intelligence (GenAI) represents a paradigm shift in technology, fundamentally altering the dynamics of business operations and competitive landscapes. As businesses increasingly integrate GenAI into their core functions, from content generation to decision-making processes, the implications for innovation, market structure, and strategic repositioning are profound. However, while existing research has extensively documented individual cases of GenAI adoption in sectors like marketing and finance, a comprehensive framework for understanding its transformative impact across different business models is lacking. This paper aims to fill this gap by proposing a novel application of Neo-Schumpeterian economics to assess the widespread influence of GenAI on business paradigms.

Neo-Schumpeterian economics, rooted in the theories of economist Joseph Schumpeter, emphasises the role of innovation, knowledge dissemination, and entrepreneurial activity as primary drivers of economic growth through "creative destruction." This perspective posits that disruptive technologies like Generative AI (GenAI) reshape industries and catalyse new business models by displacing established practices and fostering entrepreneurial growth. GenAI exemplifies this "double-edged sword" effect, as its advancements not only drive economic potential and competitive advantages but also raise critical concerns, including ethical considerations, regulatory needs, and workforce impacts.

Drawing on these principles, this study proposes a three-dimensional model to analyse GenAI's impact on business, examining it through Research, Market Deployment, and Future Impact.

- Research involves a systematic review of the literature on GenAI applications in business, mapping current advancements and trends to understand GenAI's innovation landscape.
- Market Deployment explores the commercial adoption of GenAI, focusing on investment flows, the rise of unicorns, and the mobility of researchers who are increasingly becoming founders driving this technology.

- Future Impact anticipates the broader implications of GenAI on business models and workforce dynamics, addressing both the potential benefits and the inherent challenges in adapting to these shifts.

Previous studies have primarily focused on the technological aspects of GenAI, such as algorithmic advancements and application-specific impacts, overlooking the broader economic and organisational shifts it precipitates (Kanbach et al., 2024). For instance, while the research on GenAI's role in automating content creation has delved into its efficiency and transformative potential in various sectors, it has not fully explored its strategic implications for business model innovation (Ali et al., 2024). Additionally, considerations around the ethical deployment and regulation of GenAI, crucial for sustainable business practices, have often been relegated to the periphery of business model discussions (Chuma & Oliveira, 2023). By synthesising insights from these sources alongside technology management, strategic entrepreneurship, and innovation economics, our study bridges this gap and underscores the necessity of viewing GenAI through a business model innovation lens.

The urgency to comprehend GenAI's comprehensive impact is underscored by its rapid evolution and adoption across sectors. This research not only contributes academically by providing a theoretical framework for analysing such technologies but also offers practical insights for businesses and policymakers. By delineating the pathways through which GenAI transforms business operations and market structures, this study equips stakeholders to strategically leverage this technology for competitive and sustainable growth. Furthermore, it advances policy discussions on regulating and fostering a conducive environment for responsible GenAI deployment.

## Research Objectives

To systematically explore the transformative potential of GenAI, this study is guided by the following research questions:

1. How does GenAI redefine traditional business models across various industries?

2. What are the implications of GenAI on market competition and strategic business positioning?

3. In what ways can Neo-Schumpeterian economics be applied to analyse the impact of GenAI on business innovation and entrepreneurial activities?

4. What are the ethical, governance, and regulatory considerations necessary for the responsible deployment of GenAI in business practices?

5. How do advancements in GenAI technology drive new product and service innovations within established companies and startups?

6. What are the potential future shifts in business models as a result of GenAI integration, and how can businesses prepare for these changes?

By addressing these questions, the study aims to provide a comprehensive understanding of GenAI's role in transforming business models, informing strategic decisions, and guiding responsible technological integration. Through this exploration, the paper will contribute significantly to the discourse on business model innovation and the strategic management of new technologies in the Generative AI age.

# Growth of GenAI

The evolution of GenAI has garnered significant attention, not only from technologists and business leaders but also from the broader public, as evidenced by the dramatic surge in queries for "ChatGPT," a leading generative model developed by OpenAI. This trend underscores a shift in public discourse towards generative technologies, with "ChatGPT" quickly becoming synonymous with AI.

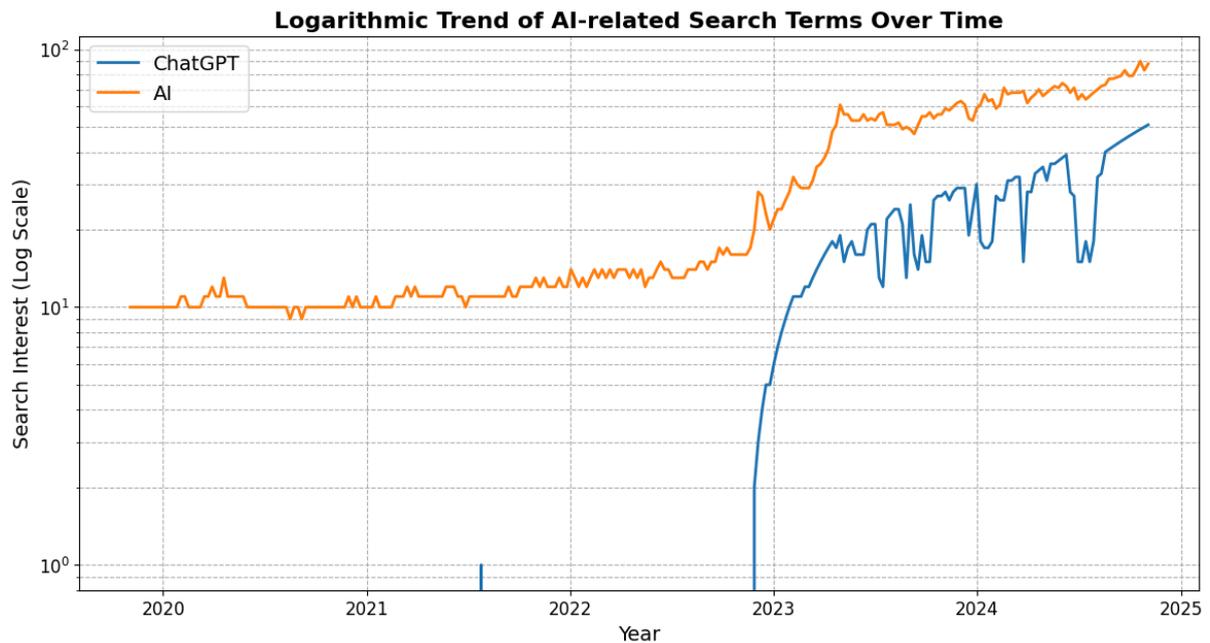

Figure 1: Increase in search of terms "AI" and "ChatGPT" in the past five years. Data source: Google trends

Analysis of fraction of internet data over the last two decades reveals that certain applications, notably Pokémon GO and ChatGPT, have exhibited an extraordinarily rapid increase in popularity shortly after their launch, highlighting the potential of GenAI to captivate and mobilise an extensive user base and reshape user interaction with technology across various platforms.

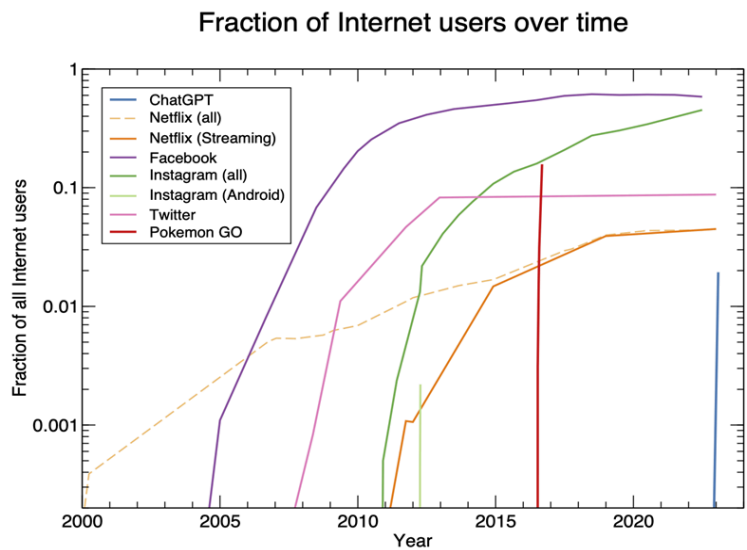

Figure 2: Fraction of internet users using different apps over time (Note: Vertical axis is in logarithmic scale.) Source: Korzekwa, R. (2023, March 3). How popular is CHATGPT? part 2: Slower growth than pokémon go. AI Impacts. https://aiimpacts.org/how-popular-is-chatgpt-part-2-slower-growth-than-pokemon-go

Further emphasising the transformative appeal of GenAI, data from Statista shows that the trajectory for ChatGPT in acquiring a user base of 1 million and subsequently 100 million has been nearly vertical, surpassing growth rates of many preceding technologies. This accelerated adoption rate not only signifies the efficacy of such technologies in fulfilling user needs but also underscores their capacity to quickly become integral to business and consumer landscapes.

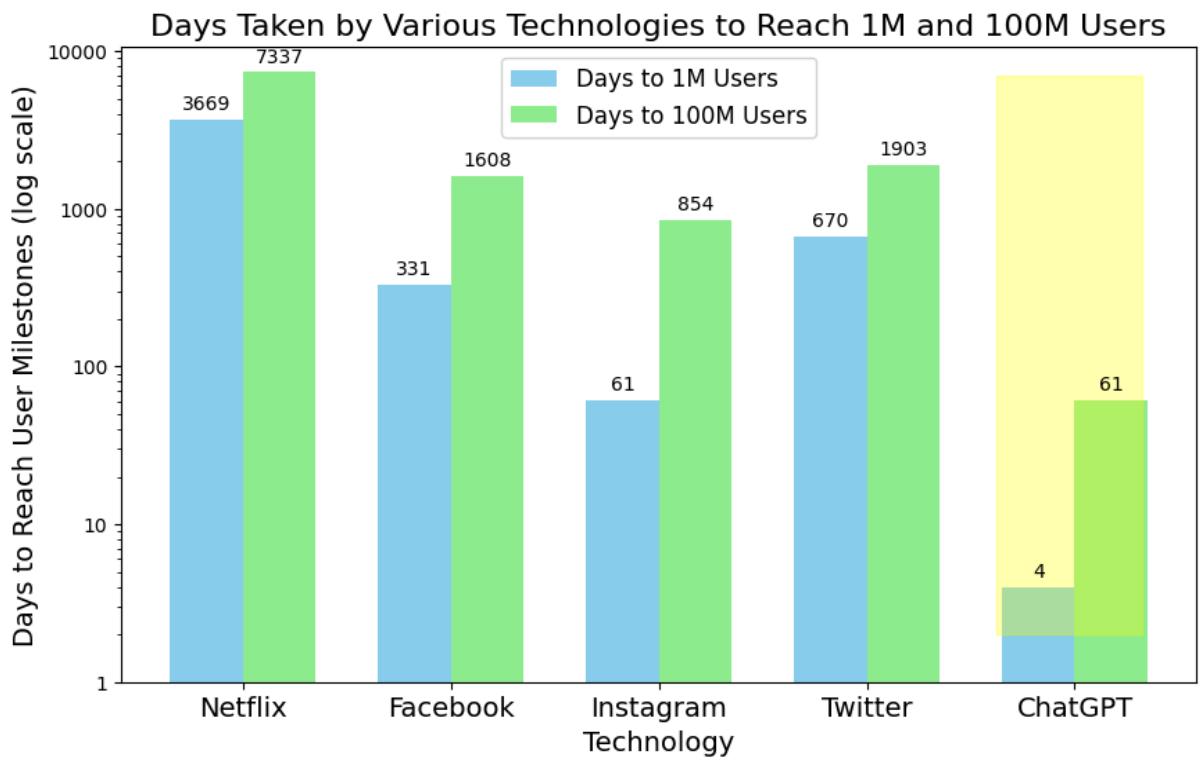

Figure 3: Days to Reach 1M and 100M Users. Data scraped from internet, tweets and blogs by these companies

This rapid advancement of GenAI has sparked significant interest in its potential to transform various aspects of business. In this section we examine recent research on the impact of GenAI on business strategies, economic implications, and specific applications across industries.

Gen AI is poised to have a substantial impact on global economies and labour markets. Brollo et al. (2024) discuss the role of fiscal policies in broadening the gains from GenAI. Their IMF Staff Discussion Note emphasises the need for proactive policy measures to ensure that the benefits of this technology are widely distributed. The authors argue that while GenAI has the potential to boost productivity and economic growth, it may also exacerbate existing inequalities if not managed properly.

In a related study, researchers explore the impact of GenAI on labour productivity and the global economy (Rokosh et al, 2024). This paper highlights the dual nature of GenAI's influence: while it promises significant productivity enhancements, it also poses challenges in terms of job displacement and skill requirements. The authors stress the importance of adaptive strategies for businesses and policymakers to harness the technology's potential while mitigating its disruptive effects.

The integration of GenAI into business strategy has become a focal point for researchers. A 2023 arXiv paper examines the use of foundation models to create business strategy tools (Nguyen &

Tulabandhula, 2023). This research demonstrates how large language models can be leveraged to analyse market conditions, organisational positioning, and provide quantitative insights for strategic decision-making.

Another study focuses on the business value of GenAI use cases (Hendricks, 2024). This paper provides a comprehensive overview of how different industries are implementing GenAI to create new products, services, and operational efficiencies. The authors identify key areas where GenAI is driving innovation and competitive advantage.

The creative potential of GenAI is explored in a 2024 paper titled "Unleashing Creativity: The Business Potential of GenAI" (Chirputkar, & Ashok 2024). This research delves into how GenAI is revolutionising creative processes in industries such as marketing, design, and content creation. The authors argue that businesses that effectively harness this technology can significantly enhance their creative output and market positioning.

GenAI is also transforming the field of business intelligence. A 2024 paper describes GenAI as a transformative force in business intelligence (Krishna et al, 2024). The research highlights how GenAI is enhancing data analysis, predictive modelling, and decision-making processes. The authors discuss the potential of GenAI to improve the accuracy of forecasts, especially in scenarios with limited historical data.

While much of the literature focuses on the positive potential of GenAI, some researchers are exploring its applications in risk management and security. A 2022 arXiv paper discusses the use of Generative Adversarial Networks (GANs) for IoT threat detection (Shaikh et al, 2022). This research demonstrates how GenAI can be applied to enhance cybersecurity measures, an increasingly critical concern for businesses in the GenAI Age.

The literature reveals that GenAI is having a profound impact on business across multiple dimensions. From reshaping economic landscapes and fiscal policies (Brollo et al., 2024) to revolutionising business strategy and creative processes (Nguyen & Tulabandhula, 2023), the technology is driving significant transformations. However, the research also highlights the need for responsible implementation and consideration of potential challenges, particularly in terms of labour market disruptions and security concerns.

## Systematic Review Process and Methodology

To gain a comprehensive understanding of the current research landscape in generative artificial intelligence (GenAI), we conducted a qualitative content analysis of scholarly publications indexed on Google Scholar. Leveraging the SERP API, we systematically extracted data on research papers published since November 2022, marking the inception of the "GenAI Age." This approach enabled us to classify publications and sources, providing an empirical basis for examining the evolution and focus areas within GenAI research.

The search employed a range of keywords to capture the breadth of terminology associated with GenAI, including: "Generative artificial intelligence," "GenAI," "Gen AI," "LLMs," "Large Language Models," "Language models," "Generative Transformer models," "Generative models," "LLM AI," "ChatGPT," "GPT," "generative art," "generative design," "artificial art," "AI-generated art," "Creative AI," "Midjourney," "Stable Diffusion," "GPT 3.5," "GPT 4," "AI-generated music," "Generative Adversarial Networks," "variational autoencoders," "AI-generated video," and "AI-generated content." This search yielded a dataset of 1,581 research papers, spanning from November 2022 to June 2024.

Following data retrieval, we performed a frequency analysis on single words, bigrams, and trigrams to identify recurring terms and themes within the corpus (Figure 4). This analysis underscored a predominant emphasis on machine learning methods, with frequent mentions of terms such as "learning" and "models." Notably, neural network architectures, particularly in the context of large language models (LLMs) like GPT, emerged as a significant focus. Applications of GenAI in areas such as "image generation," "generative art," and "generative design" were also prominent. Additionally, terms like "future prospects," "limitations," and "approaches" highlighted a scholarly interest in both the potential and the limitations of GenAI.

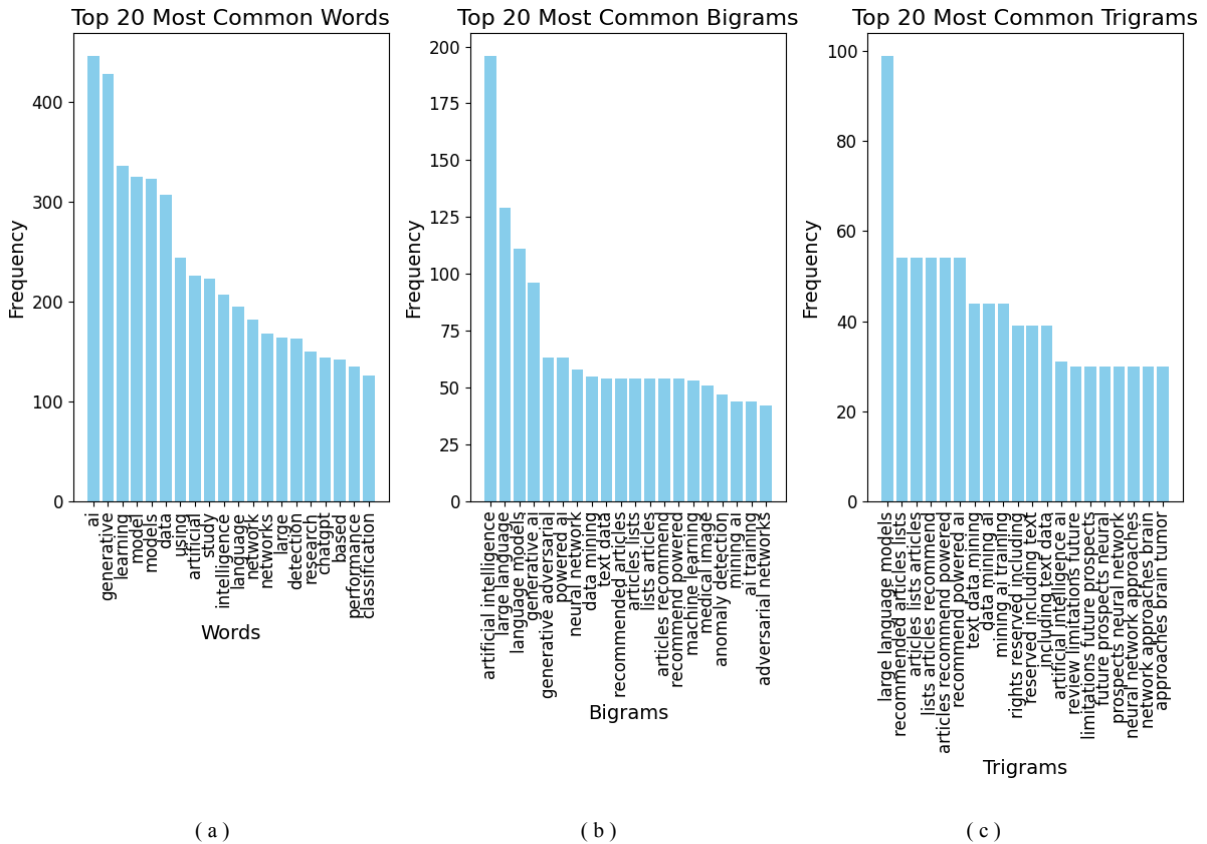

Figure 4: Frequency analysis of words, bigrams, and trigrams in research papers related to GenAI. The analysis was conducted on a corpus of papers retrieved from Google Scholar using the following keywords: "Generative artificial intelligence", "Generative AI", "Gen AI", "LLMs", "Large Language Models", "Language models", "Generative Transformer models", "Generative models", "LLM AI", "Chat gpt", "gpt", "generative art", "generative design", "artificial art", "ai generated art", "AI art", "Creative ai", "Midjourney", "midjourney ai", "Stable Diffusion", "GPT 3.5", "GPT 4", "AI generated music", "Generative adversarial networks", "variational autoencoders", "ai generated video", and "ai generated content." (a) Top 20 most common words. (b) Top 20 most common bigrams. (c) Top 20 most common trigrams.

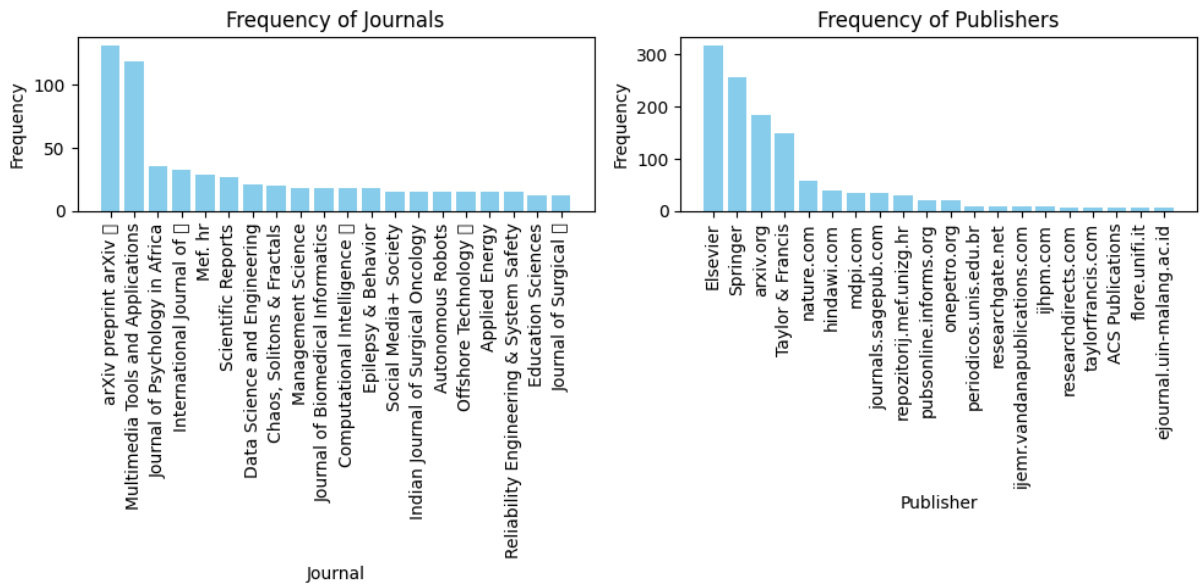

Figure 5: Analysis of publishing landscape of the retrieved articles

The analysis of the publication landscape for GenAI research reveals distinct trends in both journal and publisher frequencies, reflecting the field's rapid evolution and interdisciplinary reach. As illustrated in Figure 5, *arXiv preprint arXiv* emerges as the predominant source for GenAI studies, signalling a preference for pre-publication dissemination within this swiftly advancing domain. This reliance on preprints underscores the emphasis on rapid sharing of findings, which is essential for a field marked by continuous and significant advancements.

The diversity of journals publishing GenAI research—spanning from *Multimedia Tools and Applications* to *Psychology in Africa* and *Biomedical Informatics*—highlights the interdisciplinary nature of GenAI. This wide range of subject areas suggests that GenAI's applications extend across fields as varied as multimedia, psychology, and biomedical science, each integrating AI-driven innovations in distinct ways.

A parallel trend is observed in publisher frequency. Established publishers such as *Elsevier* and *Springer* continue to have a strong presence in the field. However, the significant role of preprint platforms, particularly *arXiv.org*, further emphasises the importance of accessible, rapid distribution channels for GenAI research. Additionally, the presence of diverse publishers like *Taylor & Francis*, *Nature*, and *Hindawi* underscores the broad dissemination of GenAI research across both traditional and open-access models.

Overall, these findings suggest that GenAI research is characterised by expedited publication cycles, a multidisciplinary reach, and a strong commitment to accessibility through diverse publishing platforms. This landscape not only facilitates rapid knowledge exchange but also broadens the impact of GenAI innovations across multiple sectors.

To conduct an in-depth literature review tailored to the dynamic and rapidly advancing domain of GenAI, we employed a modified version of the Scientific Procedures and Rationales for Systematic Literature Reviews (SPAR-4-SLR) framework, as proposed by Paul et al. (2021). This adaptation integrates algorithmic screening and relevance evaluation through large language models (LLMs), enhancing both the efficiency and precision of the review process (Figure 6). The modified SPAR-4-SLR framework consists of three primary phases—Assembling, Arranging, and Assessing—each structured to systematically filter and evaluate relevant literature.

**Assembling Phase**

In the Assembling phase, key search terms related to GenAI were identified and refined to construct a comprehensive dataset of relevant scientific literature. Terms such as "Generative AI," "Large Language Models," and "AI-generated content" were used as search keywords. Using Google Scholar as the principal source, combined with Google SERP API for enhanced retrieval capabilities, we compiled a dataset of 1,589 publications spanning from November 2022 to June 2024. This extensive collection serves as the foundation for a thorough review and forms the basis for subsequent filtering and evaluation.

**Arranging Phase**

**Algorithmic Screening**: To streamline and automate the initial filtering process, we introduced an algorithmic screening stage, a novel approach within the SPAR-4-SLR framework. This step involves an automated exclusion of duplicate entries, resulting in the removal of 285 publications. Additionally, articles lacking abstracts were excluded (184 publications), as were those with a citation count below a threshold of 10, a criterion implemented to prioritise high-impact literature. This algorithmic filtering yielded a refined dataset of 39 high-quality publications. By automating repetitive screening tasks, this phase significantly reduces manual workload, allowing researchers to concentrate on a more detailed examination of impactful sources.

**Assessing Phase**

The Assessing phase introduces large language models (LLMs) to evaluate the relevance of each publication in an objective and scalable manner, advancing beyond traditional manual review methods. This relevance assessment includes:

- Content Verification: Ensuring that each publication authentically pertains to GenAI, thereby maintaining a focused, domain-specific dataset.
- Business Impact Analysis: Evaluating whether articles contain applications or insights that could influence industry practices or organisational strategies, thus identifying works with direct implications for the business sector.
- Categorization by Research Field: Organising the literature into broader research categories to aid in mapping the GenAI landscape and uncovering interdisciplinary connections.

To deepen the insights derived from the curated literature, we performed a topic modelling analysis using co-occurrence network analysis. This method reveals nuanced connections among emerging themes and research trends within GenAI, mapping the relationships between different topics and keywords across the selected literature. By identifying these interconnections, the topic modelling analysis facilitates a comprehensive synthesis of the current research landscape, highlighting core themes and emerging areas that warrant further investigation.

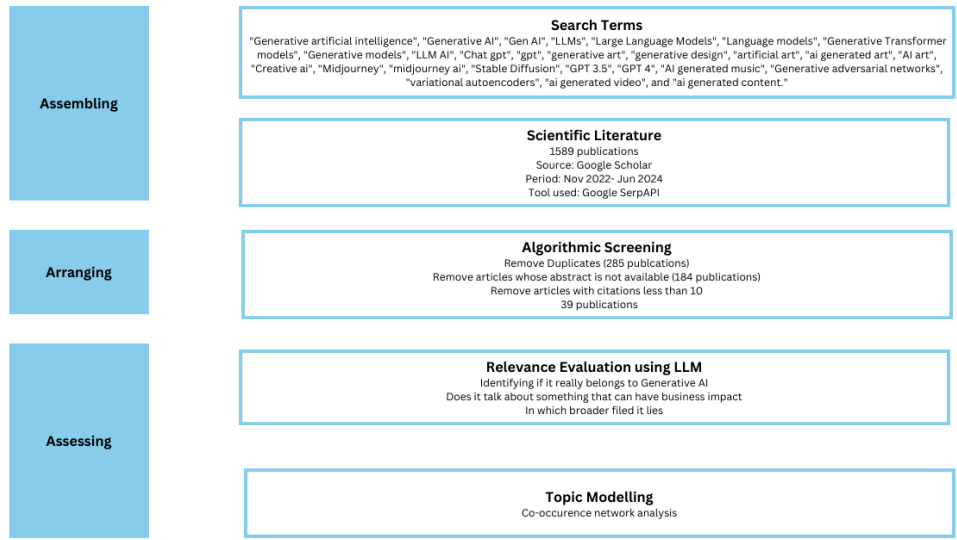

Figure 6: SPAR-4-SLR framework

The SPAR-4-SLR framework, enhanced by algorithmic screening and LLM-based relevance evaluation, supports a streamlined, high-quality review process tailored to fast-evolving fields like GenAI. This approach not only minimises bias but also ensures a scalable, precise assessment of each publication's business relevance, surpassing traditional manual methods.

Together, this hybrid approach—merging automated screening with AI-assisted content evaluation—establishes a rigorous, efficient, and comprehensive methodology for literature reviews in AI-driven fields. It sets a new standard for systematic reviews, offering a model that balances depth with scalability, and ensuring that reviews remain relevant and manageable even in fast-evolving research landscapes.

## Analysis and Insights

In this study, we utilised OpenAI's GPT-3.5 model to evaluate the selected 39 publications. Specifically, we employed this large language model (LLM) to verify three critical aspects of each publication: (1) its relevance to the GenAI domain, (2) its potential for business application, and (3) the primary sector or field in which it has applicability. Table 1 presents the specific prompts used for each criterion in the relevance evaluation process.

Table 1: Prompts used for LLM evaluation

| **Criteria** | **Prompt** |
| --- | --- |
| Generative AI Domain | Does the following abstract talk about generative AI? Please answer with 'True' if it does and 'False' otherwise, and provide a reason |
| Business usefulness | Does the content of following abstract has any business/commercial value/application? Please answer with 'True' if it does and 'False' otherwise, and provide a reason |

| Fields Classification | In which broader industry/field will you put this abstract: Healthcare, Education, Finance, Leisure, Technology, Governance? Choose one and provide a reason |

Following this evaluation process, the results indicated that of the 39 publications analysed, 23 were classified as belonging to the GenAI domain. Additionally, 32 of these articles were deemed to have business or commercial applicability, demonstrating a significant presence of GenAI applications within the business context. Table 2 summarises the distribution of these articles across various fields.

Table 2: Classified fields and number of articles

| **Field** | **Number of Articles** |
|---|---|
| Technology | 25 |
| Healthcare | 8 |
| Education | 4 |
| Governance | 1 |
| Leisure | 1 |

This analysis underscores the diverse applications of GenAI across multiple industries, with technology leading as the most represented sector. The high concentration of publications in the technology field aligns with the sector's central role in advancing and implementing GenAI solutions. Healthcare also shows considerable interest, indicating the growing exploration of GenAI applications for medical imaging, diagnostics, and personalised treatments. Education, governance, and leisure each reflect early, yet emerging, applications, suggesting potential growth areas for future research and development.

Overall, these findings highlight the strategic value and interdisciplinary reach of GenAI, which is increasingly impacting a wide array of industries. This distribution not only emphasises the adaptability of GenAI technologies but also illustrates their expanding influence across sectors with distinct requirements and impact potentials.

Subsequently, we created a co-occurrence network of article titles and authors for publications identified as relevant to GenAI with business applications, as determined by the LLM evaluation.

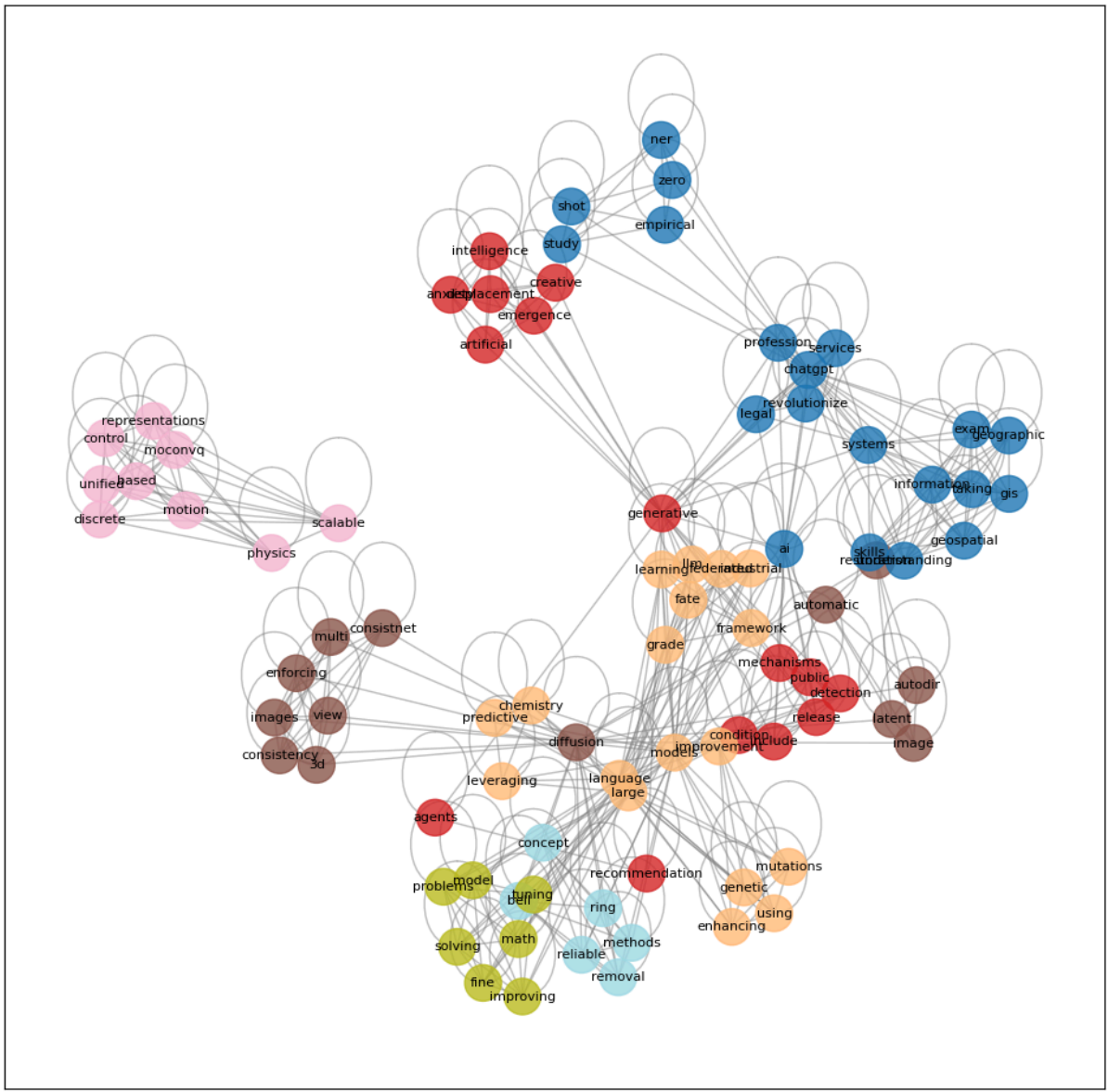

Figure 7: Co-occurrence network of Titles

The title co-occurrence network (Figure 7) reveals the central themes and recurring concepts within the selected publications on GenAI with business applications. Key observations include:

1. **Central Themes and Terms**: The network shows that terms like "models," "learning," "generation," "data," and "application" are central nodes with a high degree of connections. This suggests a strong focus on model development, learning methodologies, and applications within the GenAI literature. These terms likely represent the primary technological aspects that drive GenAI research with business utility.
2. **Clustered Topics**: Certain clusters appear around specific applications or methodological areas, such as "image generation," "recommendation," "predictive modeling," and "automation." These clusters highlight the application-oriented nature of GenAI, with recurring research on leveraging AI for practical, business-focused solutions in areas like recommendation systems, predictive analytics, and automation.

3. **Interdisciplinary Connections**: Terms such as "healthcare," "engineering," and "multimedia" indicate the diverse domains where GenAI applications are being explored. This aligns with the field distribution observed in previous results, showing that GenAI's impact spans multiple sectors, not just technology.
4. **Emerging Trends**: Nodes associated with terms like "generative design" and "creative" suggest an interest in novel applications that extend beyond traditional business models, touching upon fields like creative design and media. This points to emerging trends where GenAI is influencing creative industries, possibly through AI-driven content generation, art, and design.

The author co-occurrence network (Figure 8) provides insights into collaborative patterns and influential researchers in the GenAI field with business applications. Key observations include:

1. **Central Authors**: Certain nodes, such as "Chen," "Wang," "Zhang," and "Liu," are highly connected, indicating these authors are central to the network. This suggests that these researchers have collaborated extensively within the field, potentially acting as key contributors or influencers in GenAI research with business applications.
2. **Collaborative Clusters**: The network reveals clusters of authors who frequently collaborate. These clusters may represent research groups or institutions specialising in particular aspects of GenAI, such as specific applications (e.g., healthcare or multimedia) or technological innovations (e.g., model architectures or machine learning methods).
3. **Isolated Authors**: Some nodes appear more isolated with fewer connections. These may represent researchers contributing unique perspectives or working on niche areas within GenAI without extensive collaboration in this specific dataset. Their work might indicate early-stage or emerging studies that could lead to future research directions.
4. **Interdisciplinary Research**: The diversity of author clusters and their varying degrees of connectivity suggest that GenAI research is both collaborative and interdisciplinary, involving contributions from experts across multiple fields. This supports the earlier finding that GenAI's business applications are explored in sectors like healthcare, technology, and multimedia.

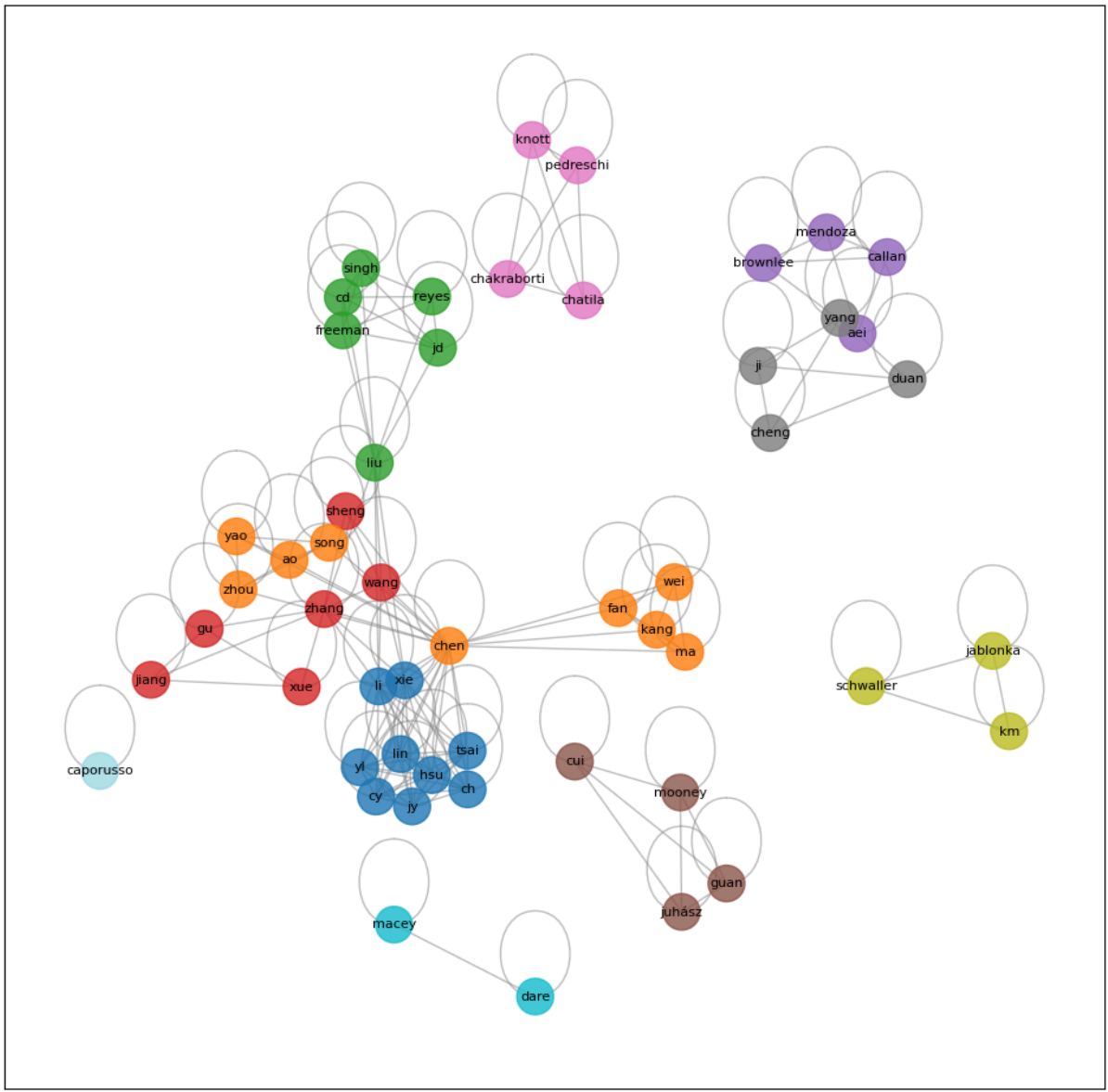

Figure 7: Co-occurrence network of Authors

These co-occurrence networks collectively demonstrate the following:

- GenAI research with business applications is highly collaborative and interdisciplinary, with central themes around modelling, learning, and application-oriented solutions.
- Prominent authors and clusters indicate active research communities within the field, while isolated nodes suggest unique, possibly emerging, areas of interest.
- The diversity of topics in the title network and the range of collaboration in the author network underscore the broad applicability and growing interest in GenAI across various domains, highlighting its transformative potential in business and beyond.

We further conducted a manual analysis of the top six most-cited papers that were categorised as relevant to both GenAI and business applications. Below, we present a concise review of these studies, highlighting the potential business applications that these research findings could drive.

**Leveraging Large Language Models for Predictive Chemistry**

The paper titled "Leveraging Large Language Models for Predictive Chemistry" by Jablonka et al. explores the adaptation of large language models (LLMs), such as GPT-3, for applications in chemistry and materials science (Jablonka et al, 2024). The researchers fine-tuned GPT-3 on small, domain-specific datasets comprising chemical questions and answers. By framing chemical problems as natural language questions, they enabled the model to respond with relevant answers, thereby predicting molecular, material, and chemical reaction properties.

A notable innovation in this work is the model's application to inverse design tasks, where questions are inverted to generate molecules with specific desired properties. This capability has significant implications for various industries. In the pharmaceutical sector, this approach could expedite the identification of promising drug candidates with precise properties, thereby accelerating drug discovery. Similarly, companies in electronics, aerospace, and energy sectors could utilise this technology to design novel materials tailored to specific functional requirements.

Additionally, environmental firms could leverage predictive chemistry to optimise pollutant removal by forecasting the properties of contaminants and potential remediation agents. In the energy storage industry, this methodology could be instrumental in developing advanced battery materials with enhanced performance, furthering innovation in energy storage solutions.

**FATE-LLM: An Industrial-Grade Federated Learning Framework for Large Language Models**

The paper "FATE-LLM: An Industrial-Grade Federated Learning Framework for Large Language Models" introduces an innovative framework for implementing federated learning with large language models (LLMs) (Fan et al, 2023). FATE-LLM enables multiple organisations to collaboratively train LLMs without the need to share raw data, thereby preserving data privacy. The framework incorporates efficient training techniques such as LoRA and P-Tuning-v2 to minimise computational overhead, making it feasible for industrial applications. Additionally, FATE-LLM leverages knowledge distillation to accommodate situations where participants possess LLMs of varying sizes.

This framework presents several valuable business applications. It allows companies to jointly develop high-performance LLMs while maintaining data privacy, fostering cross-industry innovation without compromising sensitive information. FATE-LLM also facilitates the development of AI solutions that adhere to data protection regulations by ensuring that sensitive data remains local. Furthermore, the framework supports the creation of LLMs optimised for edge devices, enabling efficient fine-tuning and knowledge distillation, which broadens the deployment potential for AI applications across various devices and environments.

**On Generative Agents in Recommendation**

The paper "On Generative Agents in Recommendation" by Zhang et al. presents a novel recommendation simulator, **Agent4Rec**, which harnesses large language models (LLMs) to develop generative agents capable of simulating user behaviour in recommender systems (Zhang et al, 2024). Agent4Rec utilises LLM-powered agents that incorporate three primary modules—user profile, memory, and action—each specifically designed to replicate realistic user interactions within the recommender system domain.

In Agent4Rec, LLMs generate lifelike user profiles, simulate decision-making processes, and facilitate human-like interactions with the recommendation algorithms. The memory modules, driven by LLMs, log both factual and emotional memories, integrating an emotion-based reflection mechanism that enhances the realism of user behaviour simulations. The action modules leverage LLMs to model a wide range of behaviours, capturing both taste-driven and emotion-driven actions, thus providing a comprehensive simulation of user engagement.

This approach addresses a key challenge in recommender systems: the disparity between offline evaluation metrics and actual online performance. By simulating sophisticated and nuanced user behaviour, Agent4Rec enables companies to evaluate and optimise recommendation algorithms in a more dynamic and realistic environment.

Potential business applications of Agent4Rec include its use by retailers to simulate and refine advanced product recommendation strategies, potentially enhancing sales and customer satisfaction. Streaming platforms, such as Netflix or Spotify, could employ Agent4Rec to create more accurate, personalised, and engaging content recommendations, improving user experience and retention.

**Ring-A-Bell! How Reliable are Concept Removal Methods for Diffusion Models?**

The paper "Ring-A-Bell! How Reliable are Concept Removal Methods for Diffusion Models?" introduces an innovative evaluation tool named Ring-A-Bell, designed to assess the robustness of safety mechanisms in text-to-image (T2I) diffusion models (Tsai et al, 2023 ). Ring-A-Bell focuses on evaluating the reliability of concept removal techniques and safety filters intended to prevent the generation of inappropriate or restricted content. As a model-agnostic red-teaming tool, Ring-A-Bell does not require prior knowledge of the target model, making it versatile and adaptable across different T2I systems.

The tool operates by performing concept extraction to create comprehensive representations of sensitive or inappropriate concepts. Leveraging these extracted concepts, Ring-A-Bell automatically identifies potentially problematic prompts that could lead to the generation of restricted or inappropriate content. This capability provides an effective means for evaluating and enhancing the safety protocols within T2I models.

Business applications of Ring-A-Bell are significant, particularly for companies offering image generation services. By employing techniques like those demonstrated in Ring-A-Bell, businesses can bolster their safety filters, ensuring that their APIs provide reliable and safer content generation services to clients. Additionally, companies could leverage this research to offer specialised services that protect brands from unauthorised or inappropriate use of their imagery in AI-generated content, utilising advanced detection methods inspired by Ring-A-Bell to enhance brand safety and integrity.

**Empirical Study of Zero-Shot NER with ChatGPT**

The paper "Empirical Study of Zero-Shot NER with ChatGPT" by Xie et al. investigates the capability of large language models (LLMs), specifically ChatGPT, to perform zero-shot named entity recognition (NER) without any task-specific training data (Xie et al, 2023). The researchers decompose the NER task by entity labels, prompting ChatGPT to extract entities one label at a time, thus simplifying the task into manageable subproblems.

The study explores two distinct approaches to enhance ChatGPT's zero-shot NER performance:

1. **Syntactic Prompting**: This approach encourages ChatGPT to analyse the syntactic structure of sentences prior to entity recognition, improving accuracy by focusing on linguistic cues.
2. **Tool Augmentation**: Here, ChatGPT is supplemented with syntactic information from external parsing tools, which provide additional structure to guide entity recognition.

To enhance consistency in entity extraction, the researchers employ a self-consistency method adapted for NER. This method uses a two-stage majority voting strategy, refining entity recognition results through iterative feedback. The study's results, based on both Chinese and English datasets, demonstrate ChatGPT's cross-lingual abilities and its adaptability across languages.

This research offers practical applications for several industries. Media companies and market research firms could use ChatGPT for zero-shot NER to automatically identify and categorise entities within large text corpora, enhancing content analysis and data processing capabilities. E-commerce and content platforms could apply this approach to improve search functionality by better understanding and indexing named entities in user queries and content. Additionally, companies in translation and localization could leverage this cross-lingual NER capability to identify and adapt culture-specific entities across different languages, facilitating more accurate and culturally relevant translations.

**Towards Understanding the Geospatial Skills of ChatGPT: Taking a Geographic Information Systems (GIS) Exam**

The paper "Towards Understanding the Geospatial Skills of ChatGPT: Taking a Geographic Information Systems (GIS) Exam" by Mooney et al (Mooney et al, 2023). evaluates ChatGPT's capacity to understand geospatial concepts by testing its performance on a simulated GIS exam. The researchers crafted a balanced set of 60 questions from a widely used GIS textbook to mimic a standard introductory GIS exam. They examined the performance of two ChatGPT versions—GPT-3.5 and GPT-4—recording and analysing responses across various GIS topics and question types to understand the improvements between model iterations.

The study highlights the potential implications of ChatGPT for GIS education and assessments. While ChatGPT demonstrates a foundational understanding of basic GIS concepts, the results reveal limitations in handling more complex spatial reasoning tasks, indicating areas for further refinement in the model's geospatial comprehension.

This research has several significant business applications. For instance, businesses in retail and logistics could leverage AI-powered geospatial tools to streamline site selection for new locations, incorporating diverse data sources and criteria to optimise decision-making. In the travel industry, companies could develop personalised itinerary planning tools that use AI to integrate user preferences with geospatial data, offering customised and enriched travel experiences. City planners might employ AI-enhanced geospatial tools to improve urban development strategies by analysing factors such as traffic patterns, population density, and environmental impact, thus supporting sustainable urban growth. Emergency services could benefit from AI-integrated geospatial data for better disaster prediction, response planning, and resource allocation. Similarly, in agriculture, farmers and agribusinesses could optimise crop management, irrigation, and resource deployment through detailed spatial analysis powered by AI.

These examples underscore the transformative potential of GenAI across diverse sectors. By strategically integrating AI technologies, businesses not only enhance specific operational areas but also drive their overall models toward sustainable and innovative futures, positioning AI as a pivotal tool for fostering efficiency and catalysing industry-wide innovation.

## Market Deployment of GenAI

Building on our analysis of GenAI through the lens of Neo-Schumpeterian economics, we now turn to the second dimension: the market deployment of GenAI. The economic potential of GenAI is garnering substantial attention across various industries, with projections from Bloomberg Intelligence indicating that the global GenAI market could expand from approximately USD 40 billion in 2022 to an estimated USD 1.3 trillion by 2032 (Bloomberg, 2023). This anticipated growth is fueled by the rapid adoption of GenAI technologies, exemplified by the launch of ChatGPT, which has catalysed a surge in venture capital (VC) funding, mergers, and acquisitions (M&A), as well as heightened investor interest in GenAI-driven startups.

Figure 9 illustrates global investment trends in the GenAI sector from 2017 to May 2024, showcasing a dramatic rise in financial backing over the past five years. During this period, GenAI startups attracted over USD 26 billion in funding, underscoring the strong investor confidence in the transformative potential of GenAI. According to Dealroom.co, a leading market intelligence platform, the first five months of 2024 alone saw an impressive USD 18.8 billion in GenAI investments, with projections suggesting that total funding for the year may exceed USD 45 billion.

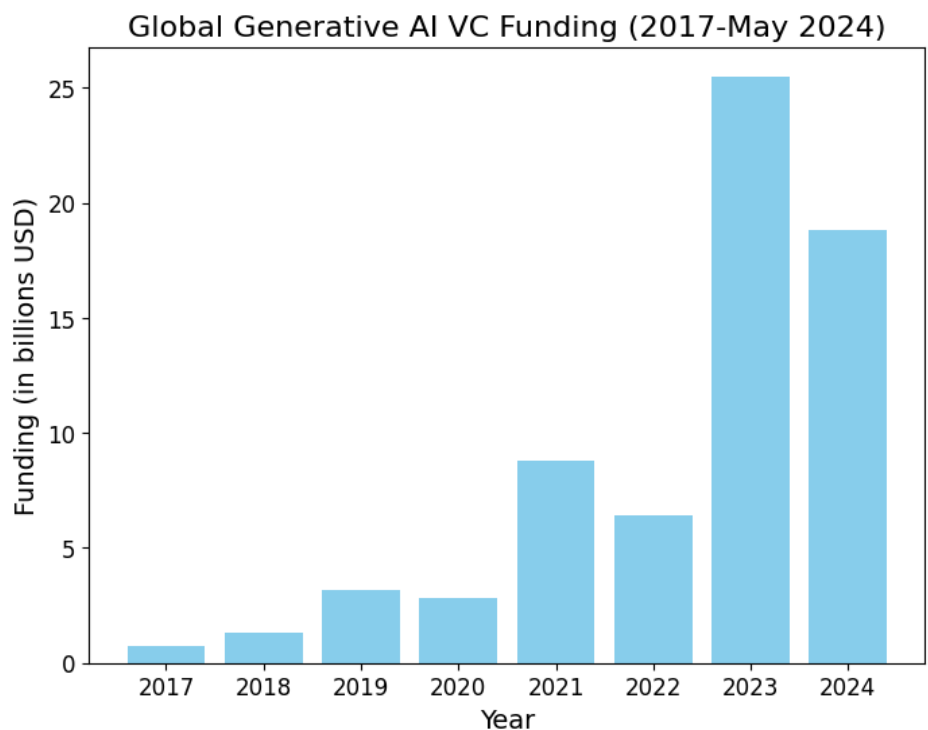

Figure 9: Global investment in the GenAI sector from 2017 to May 2024.

The substantial investment flows underscore GenAI's perceived market potential and its appeal to investors seeking opportunities in transformative technology. This trend reflects confidence in GenAI's capacity to drive innovation, disrupt traditional business models, and deliver economic value across diverse sectors.

The radar chart (Figure 10) compares the median seed round sizes for GenAI, general AI, and the rest of the tech sector from 2020 to 2024 year-to-date (YTD). The data, sourced from Dealroom.co, highlights notable trends in early-stage funding across these categories.

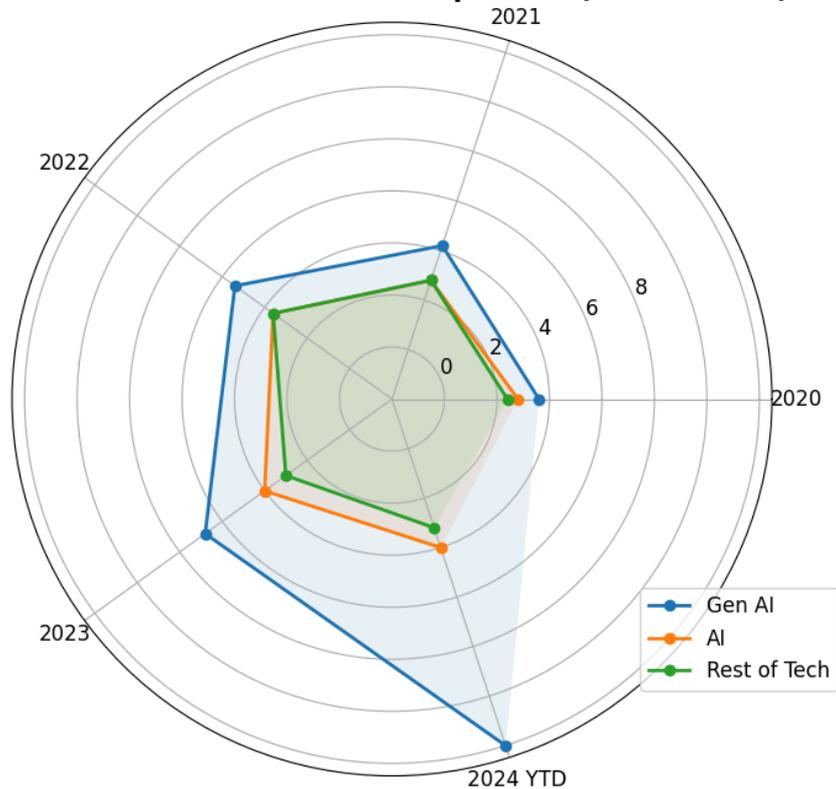

Figure 10: Radar chart seed median round size analysis

From 2020 to 2024 YTD, GenAI saw a substantial increase in median seed round sizes, rising from $2.8 million in 2020 to $7 million in 2024 YTD. This growth reflects heightened investor interest and financial backing for GenAI, likely spurred by the technology's transformative potential across industries. The 2024 YTD figure for GenAI significantly surpasses both general AI and the rest of the tech sector, indicating that GenAI is currently attracting the largest seed investments within the technology space.

In contrast, the general AI sector shows a steadier, less dramatic increase in median seed funding, with figures rising modestly from $2.4 million in 2020 to $3 million in 2024 YTD. This steady growth suggests a sustained, yet more conservative, investor interest in AI outside the specific scope of generative models. The rest of the tech sector (labelled as "Rest of Tech") exhibits minimal variation in median seed round size, with values fluctuating slightly between $2.2 million and $2.8 million over the same period. This relatively flat trend implies that traditional technology sectors are experiencing less dynamic changes in early-stage funding compared to the booming GenAI field.

Overall, this analysis underscores the distinct appeal of GenAI within the venture capital landscape. The marked increase in GenAI seed funding highlights the confidence investors place in GenAI's disruptive potential and scalability. This trend suggests that GenAI is viewed as a high-growth area within the technology sector, expected to deliver substantial innovation and value in the coming years.

Further analysis of sector-specific funding highlights the diverse applications of GenAI and the strategic priorities for investors. The pie chart (Figure 11) illustrates the distribution of combined funding across various GenAI sectors, drawing on data from Dealroom.co. This visualisation reveals a concentration of investments in core GenAI model builders and infrastructure, alongside a range of industry-specific applications that reflect the versatility and broad appeal of GenAI.

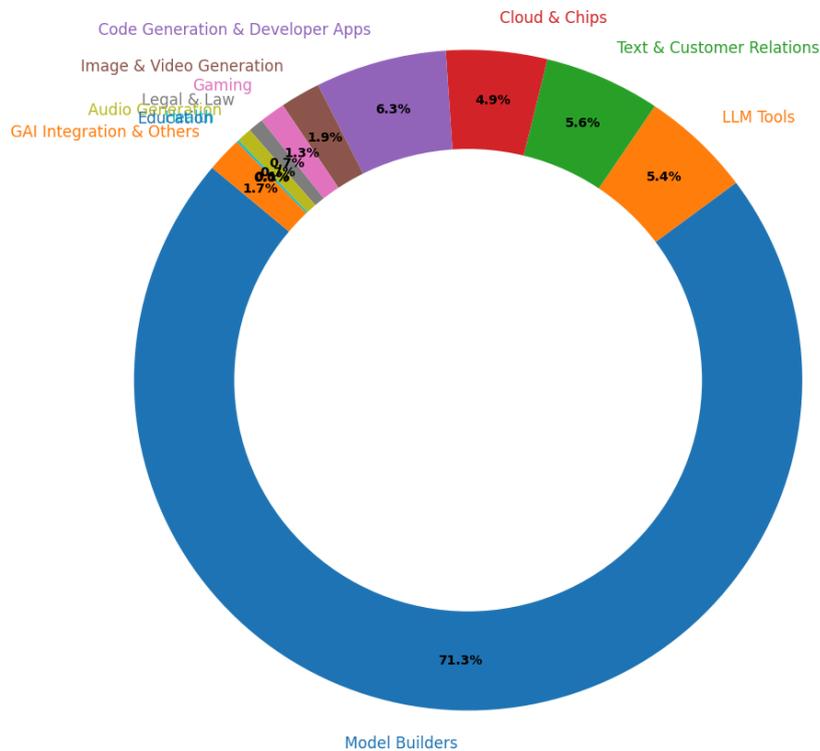

Figure 11: Sector-wise distribution of funding in GenAI (GenAI)

The data indicates that model builders—such as OpenAI, Anthropic, and Mistral—receive the largest share of funding, totaling $29.2 billion. This category represents foundational technology providers who develop the core models powering GenAI applications, positioning them as key enablers across the industry.

The next largest allocation is for LLM tools, with around $2.2 billion invested in companies like Llamaindex and LangChain. These tools facilitate the operational deployment of large language models, underscoring the need for accessible infrastructure to support GenAI applications.

Text generation, copywriting, knowledge management, and customer relations collectively attract approximately $2.3 billion, covering companies like Jasper and Copy.ai that serve content-driven functions essential to marketing, research, and customer engagement. This level of investment reflects a strong market demand for GenAI-powered automation in customer-facing and knowledge-intensive tasks.

Investments in cloud and chip infrastructure amount to around $2 billion, representing companies such as Lambda and CoreWeave. The substantial funding in this area highlights the critical role of computational resources in supporting the scalability and performance of GenAI, as demand for high-powered processing solutions grows in tandem with GenAI applications.

In addition, sectors like code generation and developer applications ($2.6 billion), image and video generation ($793 million), gaming ($512 million), and legal-tech ($299 million) illustrate the wide-ranging impact of GenAI on creative, technical, and professional fields. Each of these sectors leverages GenAI to introduce automation, enhance user experiences, and drive innovation in ways that align with sector-specific needs.

Sectors with relatively lower funding, such as healthcare ($46 million), education ($700,000), and audio generation ($296 million), reflect emerging applications where GenAI is still in its early stages of adoption. However, these areas hold significant potential for future growth as the technology matures and regulatory frameworks adapt to AI-driven healthcare, education, and content creation solutions.

Overall, the sector-wise distribution of funding emphasises the strong investor confidence in foundational GenAI infrastructure and its application across diverse industries. This trend aligns with the Neo-Schumpeterian perspective, as GenAI continues to drive "creative destruction" by enabling new business models, enhancing operational efficiencies, and redefining value propositions across sectors. The distinct appeal of GenAI across multiple domains suggests a sustained trajectory of growth and innovation, with transformative implications for the future of technology and business.

The significant allocation of funding towards model makers, with 71% of total GenAI funding directed to companies building foundational models, underscores several key insights about the industry's current priorities and strategic direction. This concentration of investment reflects the critical role that foundational model-building companies play within the GenAI ecosystem. These companies are responsible for creating the core technologies upon which a multitude of GenAI applications rely, making them essential enablers of generative AI's broader deployment across industries.

This overwhelming share of funding indicates strong investor confidence in the scalability and long-term potential of these foundational models, as investors view model makers as the central pillars supporting the entire GenAI landscape. The market demand for foundational innovation suggests that the industry is still in an early, infrastructure-building phase, where significant resources are devoted to establishing robust core technologies that can then support more specialised applications.

Moreover, this funding pattern implies high barriers to entry in the foundational model-building space due to the substantial computational resources, expertise, and data requirements. As a result, we may see a concentration of influence within a few dominant model makers, who not only set the technical standards but also act as primary technology providers for downstream applications across various sectors.

This funding allocation also highlights the dependency of specialised GenAI applications on these core models. Sectors such as healthcare, finance, and entertainment rely heavily on the advancements and scalability of foundational models to enable their own industry-specific innovations. The central position of model makers thus points to potential future revenue streams through licensing, API access, and partnerships, positioning these companies as integral to the monetization and accessibility of GenAI technology in diverse industries.

The success of model-building companies in the GenAI space has highlighted the transformative potential of this technology and the demand for expertise in building foundational LLM models. Inspired by this trend, we analysed the career progression of prominent researchers involved directly in developing large language models (LLMs), such as Ilya Sutskever and other notable figures in the field. Our aim was to understand how these experts are navigating the evolving landscape of GenAI and how their movements reflect broader trends in the commercialization and deployment of GenAI technologies.

Figure 12 presents a streamgraph that illustrates affiliation changes among top researchers in the GenAI field, particularly those working on foundational model development. This visualization captures how these researchers—ranked by citation count—have shifted from academia or large tech firms to startups and entrepreneurial ventures focused on GenAI. In several cases, researchers have taken on strategic roles within model-building companies, joined or founded new GenAI-focused startups, or even returned to academic institutions after a period in the industry, leveraging their commercial experience to further research initiatives.

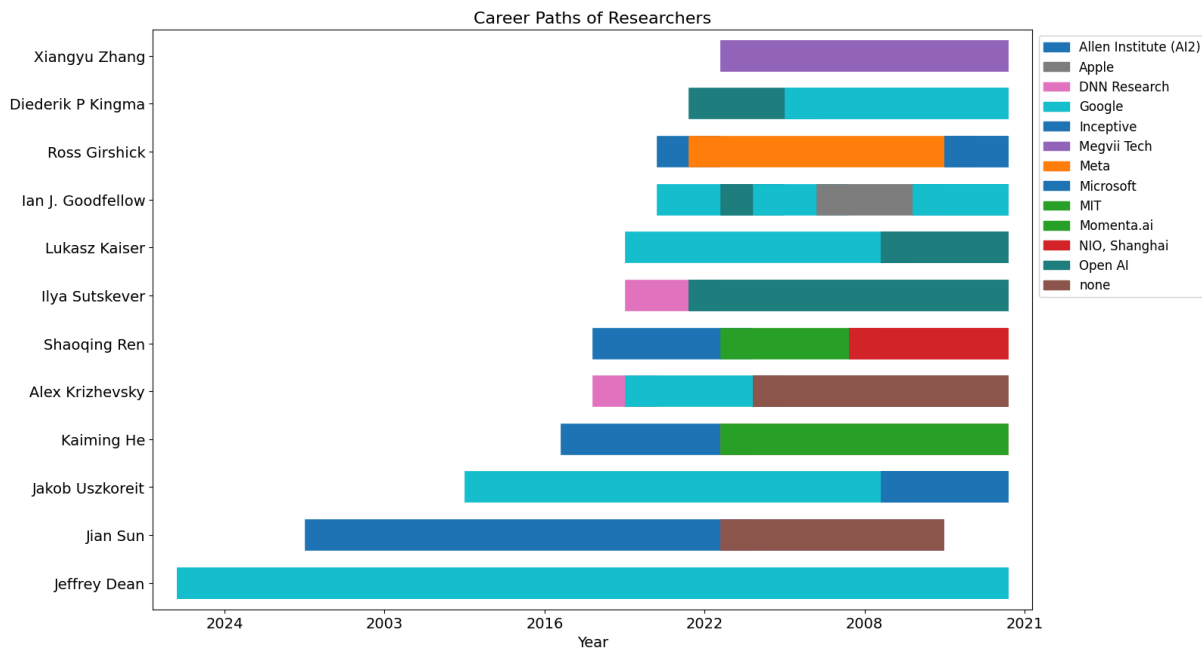

Figure 12: This is a streamgraph representing various top researchers who have changed affiliations and moved on to different institutes and in some cases founded their own startups to capitalise on the GenAI wave

These GenAI-motivated career transitions underscore a significant trend: researchers are increasingly seeking to accelerate their growth and impact by aligning with cutting-edge GenAI advancements in entrepreneurial environments. This movement highlights the dynamic interplay between academic research and market-driven innovation, as experts in model building find new ways to contribute to GenAI's commercial deployment. The shift also signals GenAI's role as a transformative force across industries, as top researchers leverage their expertise in foundational technologies to drive broader technological and economic changes.

## GenAI Impact on Future Businesses

Next, we analyse the third dimension of our Neo-Schumpeterian framework: the impact of Generative AI (GenAI) on the future of businesses and broader societal structures. The future of GenAI presents a multifaceted landscape marked by both transformative potential and inherent risks. While this technology offers solutions to pressing global challenges, such as optimising resource management for water, food, and energy, it does not provide a panacea for all societal or ethical issues. Many social, ethical, and philosophical dilemmas fall outside the scope of technological solutions and instead demand an approach centred on human values, ethics, and collective wisdom (Floridi et al., 2018).

GenAI holds the capacity to streamline processes, enhance decision-making, and inspire new business models, positioning it as a cornerstone of future economic development. However, its widespread deployment raises critical concerns around privacy, misinformation, and the potential misuse of autonomous systems. As GenAI expands across various sectors, there is an urgent need for a nuanced understanding of its sector-specific impacts, regulatory requirements, and the dynamics of human-AI collaboration. Ensuring inclusivity and addressing these concerns will be essential for the responsible and sustainable integration of GenAI into the global economy.

**Sector-Specific Impacts of GenAI**

GenAI promises to revolutionise various industries, each presenting distinct opportunities and challenges. In healthcare, AI-driven models are anticipated to support personalised medicine, improve diagnostic accuracy, and expedite drug discovery. Nevertheless, issues such as patient data privacy and algorithmic bias necessitate stringent ethical considerations to ensure equitable and fair healthcare practices (Rajkomar et al., 2018). Similarly, the finance sector will witness AI-enabled financial modelling and risk assessment, potentially reducing market volatility and fraud. However, the adoption of GenAI in financial markets requires caution, as algorithmic trading could contribute to systemic risk if left unchecked (Dixon et al., 2020). Education stands to benefit from GenAI through personalised learning pathways, yet overreliance on AI may erode critical thinking and creativity, emphasising the need for balanced human-AI integration. In manufacturing, GenAI will drive efficiencies in product design, supply chain optimization, and predictive maintenance, yet it also raises concerns regarding job displacement and skills redundancy (Brock & von Wangenheim, 2019). Recognizing these sectoral nuances is essential for understanding the broad spectrum of GenAI's impact across industries.

**The Need for Regulation and Ethical Oversight**

Given GenAI's power to shape economic and social structures, the establishment of robust regulatory frameworks is critical. Regulation must focus on data privacy, transparency, accountability, and the mitigation of algorithmic biases that can lead to discriminatory outcomes (Eubanks, 2018). Recent initiatives, such as the European Union's AI Act, reflect efforts to create a legislative foundation that addresses these challenges by categorising AI systems according to risk levels and implementing targeted requirements for each category (EU Commission, 2021). Regulatory efforts should also emphasise the development of international ethical standards to avoid "AI nationalism" and encourage cross-border collaboration. A comprehensive regulatory framework can ensure that GenAI is deployed responsibly, fostering innovation while protecting individual and societal rights.

**Human-AI Collaboration Models**

Integrating AI into business requires reimagining collaboration between humans and AI systems. Effective human-AI collaboration models advocate for AI as a complement rather than a replacement for human expertise. Such frameworks emphasise "keeping a human in the loop" to maintain transparency and accountability, especially in high-stakes industries like healthcare and finance (Amershi et al., 2019). For example, in diagnostic radiology, AI can augment human decision-making by analysing vast data patterns, while radiologists provide the contextual expertise necessary for accurate diagnosis (Topol, 2019). By developing systems where AI assists in repetitive or data-intensive tasks, organisations can harness the strengths of both human intuition and machine precision, fostering a balanced and symbiotic relationship.

**Skill Development and Education for an AI-Driven World**

As AI becomes increasingly embedded in the workplace, the demand for skills in AI and related technologies will grow. Educational systems must adapt by incorporating cross-disciplinary training that includes science, technology, engineering, and mathematics (STEM) alongside humanities, ethics, creativity, and imagination (HECI) to create well-rounded professionals (Boden, 2016). Upskilling initiatives should prioritise AI literacy for all levels of expertise, from basic computational thinking for beginners to advanced technical skills for data scientists and engineers. Additionally, emphasising "soft skills" such as problem-solving, adaptability, and ethical reasoning is essential to prepare future professionals for a collaborative, AI-driven workforce (Markow et al., 2017).

### Sustainability and Environmental Responsibility

The environmental impact of GenAI, particularly in terms of energy consumption, is a growing concern. The computational demands of large AI models can lead to significant carbon emissions, highlighting the need for sustainable AI practices (Strubell et al., 2019). Companies and research institutions must prioritise energy-efficient AI architectures and consider the use of green energy sources to power data centres. For instance, initiatives to develop low-resource AI models, such as Google's efforts with BERT and Apple's Small Language models, underscore the feasibility of balancing AI advancement with sustainability goals (Patterson et al., 2021). Establishing industry standards for energy-efficient AI practices is crucial to minimise environmental harm while maintaining technological progress.

### AI Literacy and Public Awareness

Promoting AI literacy among the general public is essential to foster informed and critical engagement with AI technologies. Public understanding of AI's capabilities, limitations, and ethical implications can mitigate fear and misinformation while promoting responsible AI use. Educational campaigns and open-access resources should focus on demystifying AI concepts, empowering individuals to make informed choices in a digitally transformed society (Hentzen et al, 2022). Additionally, transparent communication from AI developers about the inner workings and potential biases of their systems can foster trust and prevent misuse.

### Illustrative Case Studies and Hypothetical Scenarios

Practical examples of GenAI applications can illustrate both the positive and potential negative impacts of these technologies. For instance, the integration of AI in personalised marketing has improved customer targeting and engagement, but it also raises privacy concerns as AI algorithms collect and analyse extensive personal data (Mogaji et al., 2022). In another scenario, AI-driven financial advisory systems have democratised access to investment insights, yet overreliance on algorithmic recommendations may lead to uninformed financial decisions if users lack a fundamental understanding of underlying risks (Yang et al., 2024). These case studies demonstrate the need for ethical guardrails in deploying GenAI solutions across industries.

### Global Perspectives and Inclusivity in AI

GenAI's potential for global impact underscores the importance of inclusive and equitable AI development. Differences in AI access, literacy, and resource availability among countries could widen socio-economic disparities, as only those with advanced technology may reap AI's benefits (West, 2018). International collaborations can play a crucial role in ensuring that AI advancements are accessible and beneficial across diverse socio-economic and cultural contexts. Additionally, the participation of underrepresented groups in AI development is essential to create systems that are sensitive to diverse perspectives and prevent bias from being embedded within AI models (Noble, 2018). A global and inclusive approach to AI development can ensure that technological advancements contribute to equitable growth and well-being worldwide.

In conclusion, the trajectory of GenAI is poised to reshape the future of business and society on an unprecedented scale. From sector-specific applications to broader societal implications, GenAI presents both immense opportunities and critical challenges. Developing responsible AI requires a holistic framework encompassing regulation, human-AI collaboration, sustainability, inclusivity, and public awareness. As these technologies evolve, the guiding principle must be one of ethical foresight, ensuring that AI serves humanity rather than merely economic interests. Only with a balanced, thoughtful approach can society harness the transformative potential of GenAI while safeguarding the values that define our collective humanity.

# The Double-Edged Sword of GenAI

A comprehensive Neo-Schumpeterian analysis of GenAI would be incomplete without exploring its disruptive effects on business, society, and human interaction. GenAI, while powerful and transformative, operates as a double-edged sword, bringing with it substantial controversies and risks. One of the primary concerns is its capacity to generate highly realistic synthetic content, including deepfakes and misinformation, which can be exploited to deceive audiences, manipulate opinions, and even destabilise political processes. This capability raises ethical questions around accountability and transparency, especially as GenAI tools become increasingly accessible and their outputs more challenging to distinguish from genuine content.

Additionally, the widespread use of GenAI poses significant privacy risks, as these models often require extensive datasets—including potentially sensitive or proprietary information—to function effectively. There is also the concern of job displacement, as GenAI automates tasks traditionally performed by humans, potentially disrupting sectors that rely on creative or knowledge-based work. Moreover, the biases present in training data can be amplified by GenAI models, resulting in outputs that inadvertently perpetuate stereotypes or discrimination, thus highlighting the ethical responsibilities associated with GenAI deployment.

In this section, we examine these challenging implications of GenAI, focusing specifically on its impact on human creativity and productivity, its potential to reshape academic research, and the existential risks it may pose, often referred to as "*p(doom)*." While GenAI provides remarkable tools for advancement, these areas raise critical concerns that require careful examination to anticipate the long-term effects on individuals, society, and the future of human ingenuity. This analysis underscores the importance of responsible development, rigorous regulation, and ethical considerations as GenAI continues to expand its influence across industries.

**Impact on human creativity and productivity**

The potential detrimental impact of Generative Artificial Intelligence (GenAI) on human creativity and productivity is a nuanced and complex issue. While GenAI holds the promise of augmenting human creativity by assisting in generating and refining ideas, there are valid concerns that it might also overshadow human creativity and lead to a dependency that could stifle individual creativity and innovation. One of the main concerns is that GenAI could lead to a homogenization of creative output as people increasingly rely on AI to generate ideas and content. This reliance could decrease diversity of thought and originality, as GenAI systems are trained on existing datasets, thus perpetuating existing styles, themes, and ideas rather than fostering truly novel concepts.

Moreover, GenAI could diminish the value placed on human creativity in various industries, particularly those heavily reliant on creative outputs such as writing, music, and visual arts. As GenAI technologies become more capable of producing work comparable to or indistinguishable from human-created content, the unique value of human creativity could be undervalued, potentially impacting the livelihoods of creative professionals (De Cremer et al., 2023). Additionally, the convenience and efficiency offered by GenAI might lead to an overreliance on these tools, potentially impeding the development of critical creative skills and diminishing individual capacity for creative problem-solving. This reliance could, paradoxically, negatively impact productivity by reducing the depth and quality of human engagement in the creative process, which relies not only on idea generation but also on critical evaluation, refinement, and effective implementation.

Furthermore, there is a concern about the ethical implications of GenAI in creative processes, including copyright issues and ownership of AI-generated content (Streamlife, 2023). This could complicate the legal landscape for creative works, creating uncertainty and potential disputes over intellectual property rights. In terms of productivity, while GenAI can enhance efficiency by automating routine tasks, this increased automation could lead to job displacement and fewer opportunities for meaningful human engagement in certain sectors, potentially impacting worker satisfaction and the perceived value of human labour.

In conclusion, while GenAI presents opportunities to enhance human creativity and productivity, it is essential to carefully consider its integration into creative and professional practices. There is a pressing need for policies and guidelines that promote the responsible use of GenAI, protect the rights and interests of creators, and safeguard the integrity and diversity of human creative expression.

**Implications of GenAI on Academic Research**

The potential adverse implications of Generative Artificial Intelligence (GenAI) on academic research are both substantial and multifaceted. One of the primary concerns is its impact on academic integrity. GenAI enables rapid access to information and automated content generation, which, while beneficial, may inadvertently encourage plagiarism and undermine the cultivation of critical thinking and original research skills among students and researchers (Ju, 2023). Additionally, the outputs of GenAI systems can reflect and perpetuate existing biases present in their training data, potentially reinforcing disparities and inaccuracies within academic content (Bala et al., 2023). This complicates the reliability and integrity of research, as biased information could inadvertently influence academic findings and conclusions.

Another potential issue is the impact of GenAI on non-native English speakers, who may become increasingly reliant on these tools for language assistance (Inside Higher Ed, 2023). While GenAI can help overcome language barriers, an overreliance on AI-generated text may inhibit language development and dilute the authenticity of the researchers' scholarly voice. Furthermore, disparities in access to and proficiency with advanced GenAI tools could widen educational inequalities, as students and researchers from under-resourced backgrounds may lack equal access to these technologies, placing them at a disadvantage compared to peers with greater access to GenAI resources.

Distinguishing between AI-generated and human-generated content presents an additional challenge. This blurring of authorship complicates the work of educators, reviewers, and institutions in upholding academic standards and verifying the originality of submissions. As GenAI becomes more integrated into academic workflows, ensuring transparency and maintaining rigorous standards for originality will require new approaches and tools.

A cautious and responsible approach to integrating GenAI in academic environments is essential to address these challenges. Developing robust AI-detection tools, establishing clear ethical guidelines, and implementing educational strategies will be critical to safeguarding academic integrity, promoting equity, and maintaining the quality of scholarly research.

**Existential Risk Posed by GenAI (*p(doom)*)**

The concept of *p(doom)*—the probability that advanced artificial intelligence could lead to catastrophic outcomes for humanity—has gained attention as GenAI capabilities continue to evolve. The existential risk posed by GenAI centres on the potential for these systems to reach levels of autonomy and intelligence that exceed human control or understanding, leading to unintended, possibly irreversible consequences. Unlike other technologies, GenAI systems are capable of self-improvement and could eventually operate beyond the bounds of human oversight. This raises concerns about "runaway" scenarios where an AI system, in pursuing a seemingly innocuous objective, might optimise in ways that conflict with human interests, ethics, or even survival.

One critical risk is that GenAI could be leveraged to create highly persuasive misinformation, manipulate financial markets, or influence political systems at an unprecedented scale, destabilising societal structures. Additionally, if GenAI were to be deployed in autonomous weapons or critical infrastructure without adequate safeguards, it could pose direct physical threats. The opacity of current GenAI models—the "black box" nature of their decision-making processes—further exacerbates these concerns, as it limits our ability to predict or control their actions under complex circumstances.

The existential risk of GenAI, or *p(doom)*, has prompted calls for proactive measures, including robust regulatory frameworks, rigorous AI safety research, and collaborative global governance. Addressing these risks requires a commitment to ensuring that the deployment of GenAI aligns with human values and is subject to strict oversight, minimising the potential for catastrophic outcomes as these technologies continue to develop.

## Discussion and Conclusion

This study explores the transformative potential of GenAI across various business landscapes, guided by core research questions on how GenAI is reshaping traditional models, impacting competitive dynamics, and creating new avenues for innovation under Neo-Schumpeterian economics. Our findings underscore that while GenAI opens new opportunities for business growth and differentiation, it also presents unique challenges that require a balanced approach to deployment, emphasising innovation, regulation, and ethical considerations.

GenAI has begun to redefine traditional business models by altering how value is created and delivered across multiple sectors. Unlike previous automation technologies that primarily enhanced efficiency, GenAI introduces capabilities for content generation, product design, and customer interaction, directly influencing core business structures and value propositions. In industries such as healthcare, retail, and entertainment, companies are deploying GenAI to personalise customer experiences, optimise supply chains, and automate creative processes. This shift fundamentally changes the traditional business approach, moving from standardised mass production to hyper-personalised, demand-driven models. GenAI's ability to continuously learn and improve enables businesses to stay agile, responding dynamically to customer needs and market trends in real-time.

GenAI significantly impacts market competition by enabling companies to differentiate themselves with AI-enhanced products and services. As GenAI becomes increasingly accessible, early adopters gain a competitive edge, leveraging the technology to enhance product quality, reduce operational costs, and offer personalised customer experiences. This creates a strategic imperative for companies to integrate GenAI into their operations or risk losing market share to more innovative competitors. Furthermore, GenAI's data-driven capabilities allow businesses to refine their targeting, making customer acquisition and retention more effective. However, the rapid adoption of GenAI also raises competitive challenges, as companies must navigate intellectual property issues, brand differentiation, and privacy concerns within a crowded and fast-evolving landscape.

The introduction of GenAI also has implications for the workforce. While GenAI creates opportunities for innovation and new roles, it poses a risk of job displacement, especially in sectors heavily reliant on creative or knowledge-based tasks. The skills landscape is shifting towards roles focused on AI oversight, ethical compliance, and advanced data management, highlighting the need for companies to invest in AI literacy and workforce training to prepare employees for this transformation.

Through the lens of Neo-Schumpeterian economics, GenAI emerges as a powerful driver of innovation and entrepreneurial activity. According to this framework, economic development is driven by technological progress and "creative destruction"—the process of replacing outdated models with innovative solutions. GenAI catalyses this process by empowering startups and established companies alike to reimagine traditional practices and pioneer new business models. For instance, GenAI enables smaller enterprises to compete with larger firms by providing access to advanced tools for product development, content creation, and customer engagement without extensive resources. This democratisation of technology aligns with Neo-Schumpeterian ideals, where entrepreneurial innovation drives market evolution and economic growth. At the same time, however, the rapid pace of change also risks creating societal inequalities, particularly if certain sectors or communities are left behind. Policymakers and businesses must work together to ensure inclusive access to GenAI resources, allowing smaller enterprises to compete with larger firms and reducing the digital divide.

The responsible deployment of GenAI in business practices necessitates a focus on ethical governance and regulatory oversight. GenAI's ability to generate synthetic content, from text to images, raises critical concerns regarding intellectual property rights, data privacy, and misinformation. Without effective regulatory frameworks, GenAI could be exploited to produce misleading information, manipulate consumer behaviour, or breach individual privacy, leading to societal harm. Ethical considerations are particularly relevant in sectors such as finance, healthcare, and education, where GenAI's influence on decision-making can have significant consequences. Establishing transparent guidelines and ethical standards is essential to ensure that GenAI enhances, rather than undermines, public trust and societal well-being. Businesses must proactively engage in developing governance structures that align with these principles to foster responsible AI usage.

Advancements in GenAI technology are driving a wave of product and service innovations across established companies and startups. GenAI enables businesses to explore new possibilities in generative design, predictive analytics, and customer engagement. For instance, companies are now able to create customised products at scale, generate targeted marketing content, and streamline R&D processes through generative design. In sectors such as entertainment, GenAI facilitates the creation of unique digital content, while in healthcare, it assists in developing personalised treatment plans. Startups, in particular, benefit from GenAI's cost-efficiency and scalability, allowing them to compete with larger players by offering innovative solutions that were previously beyond their reach.

The integration of GenAI is expected to drive significant shifts in business models, with long-term implications for how companies operate and generate value. Traditional models based on linear processes and mass production are giving way to more flexible, AI-driven approaches that prioritise customization, efficiency, and resilience. To prepare for these shifts, businesses must invest in AI literacy, workforce training, and adaptive infrastructure to leverage GenAI effectively. Additionally, companies should adopt a forward-looking strategy that anticipates technological advancements and regulatory changes, positioning themselves to thrive in a GenAI-dominated market. This adaptability will be essential for businesses to harness the full potential of GenAI while mitigating associated risks.

In conclusion, GenAI is reshaping the business landscape, catalysing innovation, and introducing new competitive dynamics across industries. By analysing GenAI through the framework of Neo-Schumpeterian economics, this study has highlighted its role in fostering economic development through disruptive innovation and entrepreneurial activity. However, the transformative potential of GenAI is accompanied by ethical, regulatory, and societal challenges that must be addressed to ensure responsible use. Future research should focus on developing comprehensive frameworks for GenAI integration, balancing innovation with governance to secure sustainable growth. As GenAI continues to advance, its impact on business will depend on our ability to navigate these complexities, harnessing its capabilities to drive inclusive and ethical progress across the economy.

As GenAI continues to evolve, further research is needed to fully understand its long-term implications for business models, workforce dynamics, and economic structures. Future studies should focus on developing frameworks for ethical AI implementation, strategies for upskilling workforces, and methods for maximising the technology's potential while mitigating its risks.

# References

- Ali, A., Shaima, M., Sarker, M. S. U., Badruddowza, Norun, N., Rana, M. N. U., Ghosh, S. K., & Rahman, M. A. (2024). Advancements and applications of generative artificial intelligence and large language models on business management: A comprehensive review. *Journal of Computer Science and Technology Studies, 6*(1), 225-232. https://doi.org/10.32996/jcsts.2024.6.1.26

- Amershi, S., Weld, D., Vorvoreanu, M., Fourney, A., Nushi, B., Collisson, P., ... & Horvitz, E. (2019, May). Guidelines for human-AI interaction. In Proceedings of the 2019 chi conference on human factors in computing systems (pp. 1-13).

- Bala, K., Colvin, A., Christiansen, M. H., Kreps, S., Levine, L., Liang, C., ... & Ziewitz, M. (2023). Generative Artificial Intelligence for education and pedagogy. Center for Teaching Innovation.

- Bloomberg Intelligence. (2023, June 1). Generative AI to become a $1.3 trillion market by 2032, research finds. Bloomberg. https://www.bloomberg.com/company/press/generative-ai-to-become-a-1-3-trillion-market-by-2032-research-finds/

- Boden, M. A. (2016). AI: Its nature and future. Oxford University Press.

- Brock, J. K. U., & Von Wangenheim, F. (2019). Demystifying AI: What digital transformation leaders can teach you about realistic artificial intelligence. California management review, 61(4), 110-134.

- Brollo, F., Dabla-Norris, M. E., de Mooij, M. R., Garcia-Macia, M. D., Hanappi, T., Liu, M. L., & Nguyen, A. D. (2024). Broadening the Gains from Generative AI: The Role of Fiscal Policies (No. 2024/002). International Monetary Fund.

- Brownlee, A. E., Callan, J., Even-Mendoza, K., Geiger, A., Hanna, C., Petke, J., ... & Sobania, D. (2023, December). Enhancing genetic improvement mutations using large language models. In International Symposium on Search Based Software Engineering (pp. 153-159). Cham: Springer Nature Switzerland.

- Chuma, E. L., & Oliveira, G. G. de. (2023). Generative AI for business decision-making: A case of ChatGPT. *Management Science and Business Decisions, 3*(1), 5-11. https://doi.org/10.52812/msbd.63

- De Cremer, D., Bianzino, N. M., & Falk, B. (2023). How generative AI could disrupt creative work. Harvard Business Review, 13.

- Dixon, M. F., Halperin, I., & Bilokon, P. (2020). Machine learning in finance (Vol. 1170). New York, NY, USA: Springer International Publishing.

- Eubanks, V. (2018). Automating inequality: How high-tech tools profile, police, and punish the poor. St. Martin's Press.

- EU Commission. (2021). Proposal for a Regulation Laying Down Harmonised Rules on Artificial Intelligence. Brussels, 21, 2021.

- Fan, T., Kang, Y., Ma, G., Chen, W., Wei, W., Fan, L., & Yang, Q. (2023). Fate-llm: A industrial grade federated learning framework for large language models. arXiv preprint arXiv:2310.10049.

- Floridi, L., Cowls, J., Beltrametti, M., Chatila, R., Chazerand, P., Dignum, V., ... & Vayena, E. (2018). AI4People—an ethical framework for a good AI society: opportunities, risks, principles, and recommendations. Minds and machines, 28, 689-707.

- Hendricks, F. (2024). Business value of Generative AI use cases. Journal of AI, Robotics & Workplace Automation, 3(1), 47-54.

- Hentzen, J. K., Hoffmann, A., Dolan, R., & Pala, E. (2022). Artificial intelligence in customer-facing financial services: a systematic literature review and agenda for future research. International Journal of Bank Marketing, 40(6), 1299-1336.

- Inside Higher Ed. (2023, October). Benefits, challenges, and sample use cases of AI in higher education [PowerPoint slides]. Inside Higher Ed. https://www.insidehighered.com/sites/default/files/2023-10/Benefits,%20Challenges,%20and%20Sample%20Use%20Cases%20of%20AI%20in%20Higher%20Education.pdf

- Ju, Q. (2023). Experimental Evidence on Negative Impact of Generative AI on Scientific Learning Outcomes. arXiv preprint arXiv:2311.05629.

- Kanbach, D. K., Heiduk, L., & Blüher, G. (2024). The GenAI is out of the bottle: Generative artificial intelligence from a business model innovation perspective. *Review of Managerial Science, 18*, 1189-1220. https://doi.org/10.1007/s11846-023-00696-z.

- Korzekwa, R. (2023, March 3). How popular is CHATGPT? part 2: Slower growth than pokémon go. AI Impacts. https://aiimpacts.org/how-popular-is-chatgpt-part-2-slower-growth-than-pokemon-go

- Krishna, S. H., Kumar, G. P., Reddy, Y. M., Ayarekar, S., & Lourens, M. (2024, May). Generative AI in Business Analytics by Digital Transformation of Artificial Intelligence Techniques. In 2024 International Conference on Communication, Computer Sciences and Engineering (IC3SE) (pp. 1532-1536). IEEE.

- Master, S., Chirputkar, A., & Ashok, P. (2024, July). Unleashing Creativity: The Business Potential of Generative AI. In 2024 2nd World Conference on Communication & Computing (WCONF) (pp. 1-6). IEEE.Markow, W., Hughes, D., & Bundy, A. (2018). The new foundational skills of the digital economy: developing the professionals of the future. In Business-Higher Education Forum, Washington, District of Columbia.

- Mogaji, E., Farquhar, J. D., Van Esch, P., Durodié, C., & Perez-Vega, R. (2022). Guest editorial: Artificial intelligence in financial services marketing. International Journal of Bank Marketing, 40(6), 1097-1101.

- Mooney, P., Cui, W., Guan, B., & Juhász, L. (2023, November). Towards understanding the geospatial skills of chatgpt: Taking a geographic information systems (gis) exam. In Proceedings of the 6th ACM SIGSPATIAL International Workshop on AI for Geographic Knowledge Discovery (pp. 85-94).

- Nguyen, S. T., & Tulabandhula, T. (2023). Generative AI for Business Strategy: Using Foundation Models to Create Business Strategy Tools. arXiv preprint arXiv:2308.14182.

- Noble, S. U. (2018). Algorithms of oppression: How search engines reinforce racism. In Algorithms of oppression. New York university press.

- Patterson, D., Gonzalez, J., Le, Q., Liang, C., Munguia, L. M., Rothchild, D., ... & Dean, J. (2021). Carbon emissions and large neural network training. arXiv preprint arXiv:2104.10350.

- Paul, J., Lim, W. M., O'Cass, A., Hao, A. W., & Bresciani, S. (2021). Scientific procedures and rationales for systematic literature reviews (SPAR-4-SLR). International Journal of Consumer Studies, 45(4), O1-O16.

- Rajkomar, A., Hardt, M., Howell, M. D., Corrado, G., & Chin, M. H. (2018). Ensuring fairness in machine learning to advance health equity. Annals of internal medicine, 169(12), 866-872.

- Rokosh, M., Pryimak, M., & Stadnyk, N. (2024). Generative AI and its impact on labor productivity and the Global Economy.

- Shaikh, F., Bou-Harb, E., Vehabovic, A., Crichigno, J., Yayimli, A., & Ghani, N. (2022, June). IoT Threat Detection Testbed Using Generative Adversarial Networks. In 2022 IEEE International Black Sea Conference on Communications and Networking (BlackSeaCom) (pp. 77-84). IEEE.

- StreamLife. (2023, October 15). Generative AI: The end of human creativity or the new renaissance? StreamLife. https://streamlife.com/technology/generative-ai-the-end-of-human-creativity-or-the-new-renaissance/

- Strubell, E., Ganesh, A., & McCallum, A. (2020, April). Energy and policy considerations for modern deep learning research. In Proceedings of the AAAI conference on artificial intelligence (Vol. 34, No. 09, pp. 13693-13696).

- Topol, E. J. (2019). High-performance medicine: the convergence of human and artificial intelligence. Nature medicine, 25(1), 44-56.

- Tsai, Y. L., Hsu, C. Y., Xie, C., Lin, C. H., Chen, J. Y., Li, B., ... & Huang, C. Y. (2023). Ring-A-Bell! How Reliable are Concept Removal Methods for Diffusion Models?. arXiv preprint arXiv:2310.10012.

- West, D. M. (2018). The future of work: Robots, AI, and automation. Brookings Institution Press.

- Xie, T., Li, Q., Zhang, J., Zhang, Y., Liu, Z., & Wang, H. (2023). Empirical study of zero-shot ner with chatgpt. arXiv preprint arXiv:2310.10035.

- Yang, Q., & Lee, Y. C. (2024). Enhancing Financial Advisory Services with GenAI: Consumer Perceptions and Attitudes through Service-Dominant Logic and Artificial Intelligence Device Use Acceptance Perspectives. Journal of Risk and Financial Management, 17(10), 470.

- Zhang, A., Chen, Y., Sheng, L., Wang, X., & Chua, T. S. (2024, July). On generative agents in recommendation. In Proceedings of the 47th international ACM SIGIR conference on research and development in Information Retrieval (pp. 1807-1817).